\PassOptionsToPackage{inline}{enumitem} %
\documentclass[9pt,twocolumn,twoside]{osajnl}

\journal{ao} %

\setboolean{shortarticle}{false}

\setboolean{displaycopyright}{false}

\usepackage{float}
\usepackage{microtype}
\usepackage{textcomp} %
\usepackage{lineno} 
\usepackage[separate-uncertainty=true]{siunitx}
\usepackage{acro} %
\usepackage{xparse} %
\usepackage{amsfonts, graphicx, color, amsmath, mathtools}
\usepackage{array, booktabs, caption}
\usepackage{ragged2e}
\usepackage{makecell}
\usepackage{svg}
\usepackage{pgf}
\usepackage{adjustbox}
\usepackage{tikz}
\usepackage{tikzscale}
\usepackage{setspace}
\usepackage[colorlinks=true, allcolors=blue]{hyperref}
\usepackage{cleveref}

\usetikzlibrary{arrows.meta,patterns,ipe}

\acsetup{single} %

\DeclareSIPrefix\micro{\textmu}{-6}
\DeclareSIUnit\ohm{\textohm}
\DeclareSIUnit\cycle{cycles}
\DeclareSIUnit\torr{Torr}
\DeclareSIUnit\rpm{rpm}
\DeclareSIUnit\thou{thou}
\newcommand{\unsim}{\mathord{\sim}} %
\newcommand{\ungtrsim}{\mathord{\gtrsim}}

\graphicspath{{Figures/}}

\newcommand{\TE}[1]{$\text{TE}_{#1}$}

\NewDocumentCommand{\acrodef}{ m o m o }{%
	\IfNoValueTF{#2}{%
		\IfNoValueTF{#4}{%
			\DeclareAcronym{#1}{short = {#1}, long = {#3}}%
		}{%
			\DeclareAcronym{#1}{short = {#1}, long = {#3}, #4}%
		}%
	}{%
		\IfNoValueTF{#4}{%
			\DeclareAcronym{#1}{short = {#2}, long = {#3}}%
		}{%
			\DeclareAcronym{#1}{short = {#2}, long = {#3}, #4}%
		}%
	}%
}

\DeclareMathOperator*{\subjectto}{subject~to}
\DeclareMathOperator*{\maximize}{maximize}
\DeclareMathOperator*{\minimize}{minimize}

\DeclareMathOperator*{\argmin}{arg\,min}
\DeclareMathOperator*{\dB}{dB}

\DeclareAcroArticle{definite}{the}
\NewAcroCommand\dac {m}{\acrodefinite\UseAcroTemplate{first}{#1}}  %
\NewAcroCommand\Dac {m}{\acroupper\acrodefinite\UseAcroTemplate{first}{#1}} %
\NewAcroCommand\dacp  {m}{\acrodefinite\acroplural\UseAcroTemplate{first}{#1}} %
\NewAcroCommand\Dacp  {m}{\acroupper\acrodefinite\acroplural\UseAcroTemplate{first}{#1}} %
\NewAcroCommand\dacl{m}{\acrodefinite\UseAcroTemplate{long}{#1}}   %
\NewAcroCommand\Dacl{m}{\acroupper\acrodefinite\UseAcroTemplate{long}{#1}} %
\NewAcroCommand\dacf {m}{\acrodefinite\acrofull\UseAcroTemplate{first}{#1}} %
\NewAcroCommand\Dacf {m}{\acroupper\acrodefinite\acrofull\UseAcroTemplate{first}{#1}} %
\NewAcroCommand\dacfp {m}{\acrodefinite\acroplural\acrofull\UseAcroTemplate{first}{#1}} %
\NewAcroCommand\Dacfp {m}{\acroupper\acrodefinite\acroplural\acrofull\UseAcroTemplate{first}{#1}} %

\acrodef{ACT}{Atacama Cosmology Telescope}
\acrodef{CII}[{[CII]}]{ionized carbon fine structure}
\acrodef{CO}{carbon monoxide $J\rightarrow J-1$}
\acrodef{CPW}{coplanar waveguide}
\acrodef{DI}{deionized}
\acrodef{DOF}{degrees of freedom}
\acrodef{DSE}{deep silicon etch}
\acrodef{DUT}{device under test}
\acrodef{FYST}{Fred Young Submillimeter Telescope}
\acrodef{HI-VIS}{HIgh-density Vertical Integrated Spectrometers}
\acrodef{IFU}{integral field unit}
\acrodef{LCB}{lower confidence bound}[short-indefinite=an, long-indefinite=a]
\acrodef{LIM}{line intensity mapping}
\acrodef{LSS}{large scale structure}
\acrodef{MKID}{microwave kinetic inductance detector}
\acrodef{NMP}{n-methyl-2-pyrrolidone}
\acrodef{OMT}{orthomode transducer}
\acrodef{PGMEA}{propylene glycol monomethyl ether acetate}
\acrodef{PVA}{polyvinyl acetate}
\acrodef{RF}{radio frequency}
\acrodef{RIE}{reactive ion etch}
\acrodef{SO}{Simons Observatory}
\acrodef{SOI}{silicon-on-insulator}
\acrodef{SEM}{scanning electron microscope}
\acrodef{SPT}{South Pole Telescope}
\acrodef{SPT-SLIM}{South Pole Telescope Shirokoff Line Intensity Mapper}
\acrodef{SWIR}{short-wavelength infrared}
\acrodef{TE}{transverse electric}
\acrodef{TEM}{transverse electromagnetic}
\acrodef{TIFUUN}{Terahertz Integral Field Unit with Universal Nanotechnology}
\acrodef{VNA}{vector network analyzer}
\acrodef{WG}{waveguide}
\acrodef{WG-CPW}[WG--CPW]{waveguide to coplanar waveguide}[uselist={WG,CPW}, post={\acuse{WG,CPW}}] %

\makeatletter %
\newcommand{\currentfontsize}{\f@size pt} %
\makeatother

\AtBeginDocument{%
	\newwrite\widthfile
	\immediate\openout\widthfile=latex_widths.txt
	\immediate\write\widthfile{columnwidth=\the\columnwidth}
	\immediate\write\widthfile{textwidth=\the\textwidth}
	\immediate\write\widthfile{linewidth=\the\linewidth}
	\immediate\write\widthfile{fontsize=\currentfontsize pt}
	\immediate\closeout\widthfile
}

\title{A 3D-Printed Millimeter-Wave Inline Waveguide-to-Coplanar-Waveguide Transition to Enable Dense Spectrometer Arrays for Intensity Mapping Surveys} %

\author[1,2,*]{Austin Stover}
\author[3,4]{Juliang Li}
\author[5,6]{Peter Sharpe}
\author[2,3]{Morgana Iacocca}
\author[1,2]{Audrey Scott}
\author[2,3,7]{Jessica Zebrowski} %
\author[1,2,3,7,8,9]{Jeff McMahon} %

\affil[1]{Dept. of Physics, University of Chicago, 5801 S Ellis Ave, Chicago, IL 60637, USA}
\affil[2]{Kavli Institute for Cosmological Physics, University of Chicago, 5640 South Ellis Avenue, Chicago, IL 60637, USA}
\affil[3]{Dept. of Astrophysics, University of Chicago, 5801 S Ellis Ave, Chicago, IL 60637, USA}
\affil[4]{Argonne National Laboratory, 9700 S Cass Ave, Lemont, IL 60439, USA} 
\affil[5]{Dept. of Aeronautics and Astronautics, Massachusetts Institute of Technology, 77 Massachusetts Ave., Cambridge, MA 02142, USA}
\affil[6]{NVIDIA Corporation, 2788 San Tomas Expy, Santa Clara, CA 95051, USA}
\affil[7]{Fermi National Accelerator Laboratory, Fermilab P.O. Box 500, Batavia, IL 60510, USA}
\affil[8]{Enrico Fermi Institute, University of Chicago, 933 E. 56th St., Chicago, IL 60637, USA}
\affil[9]{NSF-Simons AI Institute for the Sky, 875 N. Michigan Ave., Suite 3500, Chicago, IL 60611, USA}
\affil[*]{Corresponding author: stovera@uchicago.edu}

\begin{abstract}
	We present a 3D-printed millimeter-wave, octave-bandwidth, in-line waveguide-to-coplanar-waveguide transition designed to enable focal planes with dense arrays of on-chip spectrometers. These arrays will enable compelling surveys of the large-scale structure of the universe through millimeter-wave intensity mapping. The transition consists of a four-step ridge-waveguide transformer that couples light from a rectangular waveguide onto a coplanar waveguide via an electrical connection made with indium bump bonds. We develop a tolerance-aware optimization approach to identify high-performance transition geometries that are robust to manufacturing variations; the same formulation can be applied to other tolerance-sensitive design problems. We also describe the implementation of a custom apparatus and procedure for bump-bonding a silicon chip to a metallized 3D-printed component. We detail the fabrication of the \acl{CPW} chip and three-dimensional waveguide structure, simulations and metrology of a test device, and room temperature reflectance measurements of this device. The room temperature metrology and reflection measurements are consistent with a model that predicts a coupling efficiency of $\mathbf{\unsim 95\%}$ at cryogenic temperatures in the \qtyrange{85}{170}{\GHz} frequency range.

\end{abstract}

\begin{document}
	
	\maketitle
	
	\section{Introduction}

	\begin{figure*}[t]
		\centering
		\includegraphics[width=0.9\linewidth]{IFU.tikz}
		\caption{A \acf{HI-VIS} \acl{IFU} (left) for mm-wave \acl{LIM} and a single feedhorn (right) with the \acl{OMT} made transparent and the waveguide cap removed. Each feedhorn couples incident light to an \acl{OMT}, which separates the two polarization states and routes each one to an in-line \acl{WG-CPW} transition as described in this paper. This transition couples the optical power onto on-chip spectrometers that lie parallel to the incident light, allowing the spectrometers associated with a single feedhorn to fit within its footprint. The transition therefore enables dense packing of spectrometers within a focal plane.}
		\label{fig:IFU}
	\end{figure*}
	
	Measurements of the distribution of matter on cosmological scales are key to understanding the formation, composition, and physics of our universe. Spectroscopic galaxy surveys have placed tight constraints on dark energy, the sum of neutrino masses, and primordial non-Gaussianity by building high-precision maps of \dac{LSS} over the redshift range $0 \lesssim z \lesssim 3$~\cite{collaborationDESIDR2Results2025}. Millimeter-wave \ac{LIM} surveys have the potential to complement and expand on these measurements in higher, relatively unexplored redshifts, efficiently probing the \ac{LSS} over large scales and providing access to more information about the Universe~\cite{karkareSnowmass2021Cosmic2022}.
	
	\ac{LIM} spectroscopically measures the integrated emission from spectral lines in galaxies and the intergalactic medium, including the \ac{CO} and \ac{CII} emission lines at millimeter wavelengths, enabling the tomographic reconstruction of large scale structure at higher redshifts ($2 \lesssim z \lesssim 10$). \ac{LIM} surveys represent a powerful new approach to precision cosmology, but aggregate line emission is faint, making sensitivity a central challenge. 
	
	Research in \ac{LIM} spectrometers is therefore focused on maximizing sensitivity.  A key approach is to push to higher densities of detectors within a fixed focal-plane area. Several current and planned \ac{LIM} experiments, such as \dac{SPT-SLIM}~\cite{karkareSPTSLIMLineIntensity2022} and \dac{TIFUUN}~\cite{rybakTHzIFUsDESHIMA2024}, use on-chip spectrometers that perform simultaneous readout of hundreds of frequency channels organized into multiple spectrometers within a monolithic focal plane. These spectrometer arrays are compact and can provide higher instantaneous sensitivity per unit focal plane area than other \ac{IFU} implementations~\cite{betherminCONCERTOHighfidelitySimulation2022,critesTIMEPilotIntensityMapping2014,vieiraTerahertzIntensityMapper2019}. However, a substantial fraction of the focal plane area of these arrays is devoted to the spectrometers, which are large compared to each optical coupling element. An order-of-magnitude improvement in focal plane packing density may be obtained by arranging each spectrometer to fit within the footprint of its optical coupling element. This paper demonstrates a key technology needed to realize this goal: an in-line \ac{WG-CPW} transition. With this transition, the on-chip spectrometer can be rotated perpendicular to the focal plane, such that incident light approaches the edge of the spectrometer chip. Light entering a feedhorn antenna is then coupled in-line from a \ac{WG} onto a spectrometer that fits within its feedhorn's footprint. %
	
	\Cref{fig:IFU} shows a preliminary concept for a densely packed spectrometer array for mm-wave intensity mapping enabled by the in-line transition presented in this paper. This design packs 384 spectrometers into an \qty{18.9}{\mm} diameter hexagonal focal plane, demonstrating that an order of magnitude improvement in packing density is possible with this approach. Each optical coupling element consists of a feed-horn~\cite{simonFeedhornDevelopmentScalability2018,simonDesignCharacterizationWideband2016} which couples light from the telescope optics through a circular waveguide to an \ac{OMT}. The \ac{OMT}~\cite{shenCompactWBandSwan2020,gomez-torrentCompactSiliconMicromachinedWideband2019} separates the two incident linear polarization states and outputs each polarization through a rectangular waveguide to an in-line \ac{WG-CPW} transition as described in this paper. The \ac{WG-CPW} transition then couples the signal onto a feedline for \acp{MKID} on a filterbank spectrometer-on-a-chip, with readout through wire bonds to SMP connectors~\cite{stoverVerticalIntegratedSpectrometer2024,stoverHighdensityVerticalIntegrated2025}. This architecture, which we call \dacf{HI-VIS} \ac{IFU}, provides a path to realize cameras containing thousands of spectrometers on modern telescopes like \dac{FYST}, \ac{SO} telescopes, or \dac{SPT}.  Such instruments would provide the mapping speed needed to survey large cosmological volumes and achieve the scientific potential of mm-wave line intensity mapping~\cite{karkareSnowmass2021Cosmic2022}.
	
	\Cref{sec:design} presents the design of the \ac{WG-CPW} transition and describes its implementation in a test device. \Cref{sec:opt} describes the optimization and tolerancing methodology to choose a nominal design point for this transition, and \cref{sec:fab} details the fabrication of the three-dimensional and on-chip components of the transition. \Cref{sec:bond} describes the interconnection of these components through bump bonding and the development of a custom bump-bonder that can perform this function. Finally, \cref{sec:meas} reports reflectivity measurements of the test device and compares these measurements to simulations to characterize device performance. The sources of loss are accounted for, both in the measurement environment and extrapolated to cryogenic temperatures, and this combination of measurement and modeling is used to estimate the transmission of a transition fabricated for cryogenic applications. \Cref{sec:conclusion} concludes.

	\section{Design}
	\label{sec:design}
	The in-line \ac{WG-CPW} transition is designed to provide an octave bandwidth: \qtyrange{85}{170}{\GHz}. The waveguide side of the transformer is adapted from the in-line waveguide-to-coaxial transition presented by Cano and Mediavilla~\cite{canoOctaveBandwidthInline2017}, retaining a stepped-ridge impedance transformer and electrically small coupling gap before the backshort. In the present design, the waveguide is scaled for higher frequencies and the coaxial output is replaced by a \ac{CPW}, requiring a new planar output geometry, reoptimization of the waveguide structure, and a different interconnect.
	
\begin{figure}[t]
		\centering
		\includegraphics[width=\linewidth]{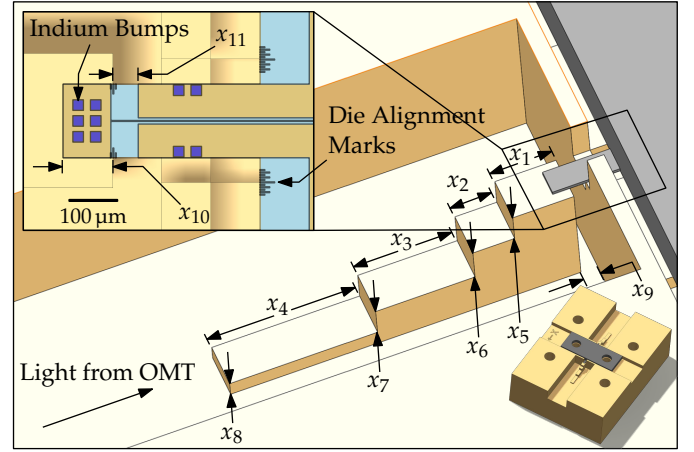}
		\caption{Diagram of the transition, including the \acl{CPW} chip as viewed from the handle side, with detail of the cantilever and a render of the overall back-to-back transition test device inset. The \acl{WG} cap is omitted for clarity. Note that the detail of the cantilever shows bump bonds and traces deposited onto the device side (the underside) of the cantilever. Dark blue denotes indium bumps on the underside, gold denotes traces on the underside, and light blue denotes silicon. The underlying \acl{WG} structure is yellow. Dimensions are given in \cref{tab:params}.} 
		\label{fig:Diagram}
	\end{figure}

	The transition has two main components: a 3D waveguide with the stepped ridge transformer (the waveguide structure) and a 2D \ac{CPW} fabricated on a chip (the \ac{CPW} chip). Polarized light enters the transition through the rectangular waveguide as a \acl{TE} \TE{10} mode and ascends the 4-step ridge waveguide impedance transformer (\cref{fig:Diagram}). This step structure both matches the impedance of the waveguide to that of the \ac{CPW} and is optimized such that reflected modes in the transition destructively interfere. A benefit of using a ridge waveguide is that it has frequency cutoffs that allow for extended single-mode performance, enabling wide bandwidth. The steps also reshape the propagating field to concentrate power at the ridge, matching more closely the field configuration of the \ac{CPW} and its dominant mode. An electrically small gap between the back of the last step and the backshort acts as a controlled reactance that transforms the waveguide \ac{TE} mode to the quasi-\ac{TEM} \ac{CPW} mode~\cite{canoOctaveBandwidthInline2017}. This gap must be small to ensure that it suppresses any additional unwanted resonant modes that would reduce the passband width. A cantilevered silicon probe carrying the \ac{CPW} on the chip extends into the waveguide from beyond the backshort. It protrudes from a rectangular aperture in the backshort which extends to the top of the waveguide for assembly, but is small enough that radiation at the relevant frequencies cannot leak out (either as a propagating mode or via tunneling). The conductor trace on this probe is galvanically connected to the top step with indium bump bonds, and each ground trace is likewise connected to the backshort. Indium is chosen for the interconnect since it remains ductile even at temperatures approaching absolute zero and can be deposited as a thick film~\cite{lucasIndiumBumpProcess2022}. A trough extrudes opposite the \ac{WG} structure from the backshort underneath the \ac{CPW} conductor trace to accommodate the portion of the \ac{CPW} mode that propagates in the air.
	
	One important design modification required by moving from the coaxial probe in~\cite{canoOctaveBandwidthInline2017} to a \ac{CPW} is that the \ac{CPW} ground traces must protrude beyond the backshort and into the waveguide a certain distance $x_{11}$~(\cref{fig:Diagram}). This ground extension is key to ensuring good performance of the transition; lossless simulations show that a ground trace extension of optimal length can decrease reflected power by around an order of magnitude with respect to the same transition without a ground trace extension.
	
	\subsection*{Test Device}
	The test device consists of two back-to-back \ac{WG-CPW} transitions connected by a \ac{CPW} of length \qty{2.6}{\milli \meter} end-to-end. The two transitions comprise a monolithic three-dimensional \ac{WG} structure bump-bonded to a die with a single \ac{CPW} as shown in the inset of \cref{fig:Diagram}. A planar metal cap is added to close each waveguide in the transition; this cap is connected to a standard waveguide flange for connections to millimeter-wave test equipment~(\cref{fig:DUT}).
	
	A channel underneath the \ac{CPW} conducting trace in between the back-to-back transitions permits the propagation of a transmission line mode. Two \qty{745}{\micro \meter}-wide raised features, known as bosses, flank both sides of this \ac{CPW} channel (as seen in \cref{fig:Confocal}); these bosses are bump bonded to the \ac{CPW} ground traces with large bond arrays (the upper right image in \cref{fig:InAndSWIR}). The bond arrays mechanically join the waveguide transition structure to the chip as well as create an air-bridge across the \ac{CPW} ground traces which suppresses odd modes. The purpose of these bosses is to limit any small warping of the 3D-printed back-to-back transition from affecting the relative orientation of the die. Deviations in the flatness of the back-to-back transition are additionally accounted for by depositing indium both onto the die and onto the back-to-back transition: a combined target thickness before bonding of  $\unsim\qty{20}{\text{\micro\meter}}$ of indium between the die and waveguide transition structure provide the allowance necessary to ensure a reliable bond; the bond is performed with a goal of $\unsim \qty{50}{\%}$ compression.
	
	During the design, it was envisioned that this test device could facilitate both reflection measurements and transmission measurements. After fabrication, it was realized that losses in the silicon device layer were large enough that transmission is exceedingly small. Dielectric losses alone~(\cref{eqn:losstan,eqn:losstaneps}) are estimated at $\ungtrsim\qty{30}{\decibel}$.  Losses associated with the gold conductive traces make this substantially higher.   High-quality transmission measurements would require a cryogenic testing environment and a new die fabricated on a high-resistivity (float-zone) silicon substrate with superconducting niobium conductors. This is beyond the scope of the present work.
	
	\subsection*{Implementation with Microstrip}
	The design presented here is based on a \ac{CPW} but implementation with a microstrip is possible and advantageous in some circumstances. We present one such design in \cref{app:microstrip}.
	
	\section{Optimization and Tolerancing}
	\label{sec:opt}

	\begin{figure}[t]
		\centering
		\input{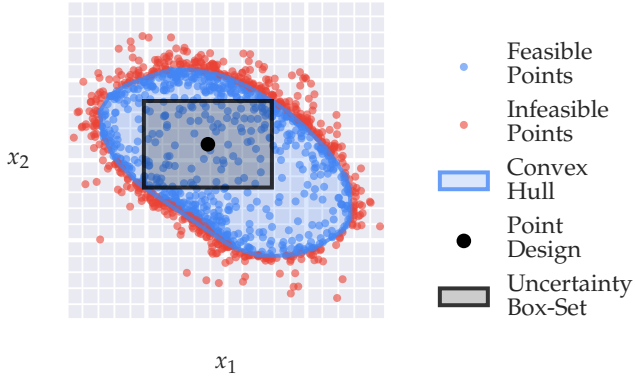} %
		\caption{Illustration of the robust optimization procedure~(\cref{eqn:optalg}) for a toy cost function $f(x_1,x_2)$. The procedure identifies a point design $x^\star$ (black marker) for fabrication which maximizes the volume of the uncertainty box-set $\mathcal{B}(\mathbf{x},\mathbf{\Delta x})$ (black boundary). The uncertainty box-set represents a fabrication tolerance around the point design and lies within the set of feasible designs $\mathcal{F}$: designs that satisfy a desired performance constraint $f(x_1,x_2) \leq f_0$ (blue markers). Designs that do not satisfy the feasibility constraint are marked in red. The performance level set---the boundary of the set of feasible designs around $\argmin f(x_1,x_2)$---is approximated by the convex hull of feasible points (blue polygon) as sampled by a Bayesian optimizer.}
		\label{fig:optimization}
	\end{figure}
	
	Optimization of the transition geometry was carried out using a novel formulation to both achieve good transmission across the passband and to minimize the sensitivity of the final design to manufacturing tolerances. Simulations of candidate designs were performed using a third-party mesh-based electromagnetic solver\footnote{Ansys HFSS 2022 R2 (Ansys, Inc., Canonsburg, PA, USA).} to analyze their performance. However, each black-box simulation is computationally expensive, limiting the ability to explore the 11-dimensional design space (\cref{fig:Diagram}) ~\cite{sharpeAcceleratingPracticalEngineering2024,sharpeAeroSandboxDifferentiableFramework2021}. In addition, cost functions~(\cref{eqn:cost}) characterizing insertion loss in this design space were found to be highly nonconvex (i.e. many local optima), so manual tuning combined with a gradient descent method is similarly inefficient. We therefore use Bayesian optimization~\cite{nogueiraBayesianOptimizationOpen2014,gardnerBayesianOptimizationInequality2014} to efficiently sample the design space and perform the optimization in an automated manner. Bayesian optimization models the cost function as a Gaussian process, a model which may be evaluated relatively cheaply and often compared to the electromagnetic simulation. This model is used to select the most promising candidate for the next evaluation, updating its posterior belief after evaluation to select the next candidate, so it is able to explore a large, highly nonconvex space relatively quickly.
	
	\subsection*{Approach}
	Typically, waveguide transition design optimization is performed without taking into account manufacturing tolerances in the optimization methodology, but the performance of this mm-wave transition was found to be highly sensitive to deviations in the geometry on the order of $\unsim$\qty{10}{\micro\meter}, which is around or below the tolerances of several of the waveguide structure fabrication techniques that were considered~\cite{bostonmicrofabricationComparingSLADLP,kalpakjianManufacturingProcessesEngineering2022a}. Thus, considering uncertainty in the optimization process allows us to identify a design that not only exceeds a target performance threshold at its nominal design point but also exceeds this threshold for any design within manufacturing tolerances.
	
	We quantify the performance of a candidate design $\mathbf{x} \in \mathcal{X}$ (with parameters $x_1,x_2,\cdots,x_N$), where $\mathcal{X} \subseteq \mathbb{R}^N$ is the search domain, with a goal-based scalar performance cost function
	\begin{equation}
		f(\mathbf{x}) = \left\Vert\max{\left(L_{\nu_i}(\mathbf{x})-L_\text{goal},\;0\right)}\right\Vert_2
		\label{eqn:cost}
	\end{equation}
	where the frequency-dependent loss is
	\begin{equation*}
		L_{\nu_i}(\mathbf{x}) = 20\log_{10}|S_{11}(\nu_i;\mathbf{x})| \\
	\end{equation*}
	and $||\mathbf{r}||_p = \left(\frac{1}{N_\nu}\sum_{\nu_i} r_{\nu_i}^2\right)^{1/p}$ is the $L_p$ norm on the residuals $\mathbf{r}$. $L_{\nu_i}(\mathbf{x})$ measures the reflected power $|S_{11}|^2$ in decibels across the sampled frequencies $\nu_i$.\footnote{If loss is not negligible, as in simulations with lossy materials or where power may be radiated away, the power lost during transmission, $1-|S_{21}(\nu_i;\mathbf{x})|^2$, may be used in place of $|S_{11}(\nu_i;\mathbf{x})|^2$. In this paper we optimize a \ac{WG-CPW} transition with approximately lossless materials and include losses only after identifying optima in order to model the test device. The \ac{WG} to microstrip transition in \cref{app:microstrip}, on the other hand, was optimized using $1-|S_{21}(\nu_i;\mathbf{x})|^2$ as the loss.} For the \ac{WG-CPW} transition, $L_\text{goal}$ is the desired reflected power level, in decibels, across the passband. We use $p=2$ for a balance between penalty and flexibility in the loss function.
	
	Directly minimizing $f(\mathbf{x})$ (na\"ive optimization) identifies a geometry with low reflected power, but it does not determine how performance declines under manufacturing deviations. When the minimum $\mathbf{x}^\text{min} = \argmin{f(\mathbf{x})}$ lies in a narrow or anisotropic valley, dimensional errors can move the fabricated transition to a nearby geometry with substantially higher reflection. We therefore use $f(\mathbf{x})$ not as the optimization objective function but as a local performance criterion.
	
	Instead, we approach the problem with a set-based optimization formulation similar to the minimax approach: instead of finding the single best design, we find the largest axis-aligned box $\mathcal{B} \subseteq \mathcal{X}$ in the $N$-dimensional parameter space such that
	\begin{equation*}
		\mathcal{B}(\mathbf{x},\mathbf{\Delta x}) = \prod_{i=1}^N \left[ x_i - \Delta x_i,\;x_i + \Delta x_i \right], \qquad \mathbf{\Delta x} \in \mathbb{R}^N_{\geq 0}
	\end{equation*}
	where the cost function values $f(\mathbf{x}) \in \mathbb{R}$ of even the worst designs contained by that box lie within some desired performance threshold $f(\mathbf{x}) \leq f_0$. This box represents an uncertainty set: the set of designs that would be compliant with a typical per-dimension manufacturing tolerance specification. In other words, for some measure of the box size $\Phi(\mathbf{\Delta x})$, we solve
	\begin{equation}
		\begin{aligned}
			&\maximize_{\mathbf{x}, \mathbf{\Delta x}}      &   \Phi(\mathbf{\Delta x})& \\
			&\subjectto                                     & f(\mathbf{y}) \leq f_0& \qquad \forall \mathbf{y} \in \mathcal{B(\mathbf{x}, \mathbf{\Delta x})} %
		\end{aligned}
		\label{eqn:optalg}
	\end{equation}
	to determine the largest possible design tolerances that still guarantee the desired performance. $\Phi(\mathbf{\Delta x})$ represents the objective function for this robust optimization formulation, whose optimal point design, $\left(\mathbf{x^\star}, \mathbf{\Delta x^\star} \right)$, is equivalent to a manufacturing specification for a nominal design and tolerance: $x_i^\star \pm \Delta x_i^\star$.
	
	\subsection*{Implementation}
	\label{sec:opt_implementation}
	To implement \cref{eqn:optalg} we use a two-step optimization procedure as illustrated in \cref{fig:optimization}. In the first step, we identify the optimization constraint and domain. We determine a target performance threshold $f_0$ and aim to find the corresponding feasible set $\mathcal{F} = \left\{ \mathbf{x} \;\middle|\; f(\mathbf{x}) \leq f_0 \right\}$ in parameter space. To do this, we developed a script that automates execution of the third-party electromagnetic solver in order to integrate it into our custom simulation algorithm. With this script, we implement Bayesian optimization using domain reduction~\cite{standerRobustnessSimpleDomain2002} to minimize $f(\mathbf{x})$. Then we use Bayesian optimization to efficiently sample the designs near the feasibility boundary surrounding $\mathbf{x}^\text{min}$.
	
	To implement this boundary sampling, we aim to determine the feasibility boundary around $\mathbf{x}^\text{min}$. We transform the problem into the optimization of a sigmoid curve within some smaller domain $\mathcal{Z} \subset \mathcal{X}$ which covers $\mathbf{x}^\text{min}$ and where we also expect the boundary to lie:
	\begin{equation*}
		\minimize_{\mathbf{z} \in \mathcal{Z}} \quad   \tanh{\frac{f(\mathbf{z}) - f_0}{s}}
	\end{equation*}
	where $s$ is a domain scaling factor that here we set to unity. The sigmoid curve translates the feasibility region into a smooth, differentiable ``basin''. We perform the optimization with a Bayesian approach using \iac{LCB} acquisition function $a_{LCB}(\mathbf{x}; \beta) = \mu(\mathbf{x}) - \beta \sigma(\mathbf{x})$ with a large tradeoff parameter $\beta= 100$, where $\mu(\mathbf{x})$ and $\sigma(\mathbf{x})$ are the mean and marginal standard deviation functions of $f(\mathbf{x})$, respectively~\cite{srinivasInformationTheoreticRegretBounds2012,kochenderferAlgorithmsOptimization2019}. The acquisition function determines where the next evaluation occurs in the problem space---it gives the expected loss for the Gaussian random process associated with the evaluation of $f$ at $\mathbf{x}$. A relatively heavily weighted $\mu(\mathbf{x})$ prioritizes exploitation whereas weighting $\sigma(\mathbf{x})$ prioritizes exploration, so by setting $\beta \gg 1$, the optimizer prioritizes exploration near regions where the variance $\sigma^2(\mathbf{x})$ of $f(\mathbf{x})$ is high---that is, on the feasibility boundary where the sigmoid curve produces the most variance.
	
	The subset of the sampled designs within $\mathcal{F}$ is then used to construct a convex hull, the smallest convex set (or polyhedron) that encloses this subset. This convex hull approximates the performance level set $\left\{ \mathbf{x} \;\middle|\; f(\mathbf{x}) = f_0 \right\}$ and can be mathematically represented as a polytope of the form $A\mathbf{x} \leq \mathbf{b}$, where $A \in \mathbb{R}^{M \times N}$. Here, $M$ is the number of facets of the convex hull and $N$ is the number of dimensions in the design space, with $M \gg N$ typically.
	
	In the second step, we aim to maximize $\Phi(\mathbf{\Delta x}) = V(\mathcal{B})$, the volume of the axis-aligned box $\mathcal{B}$ contained within this convex hull, according to the constraint and feasible set identified in the first step. Specifically, this forms a robust optimization problem where the variables are the box's center coordinates and side lengths: $(\mathbf{x}, \mathbf{\Delta x})$. In general, this yields a box with $2^N$ vertices, all of which must be contained within the convex hull. Each vertex therefore produces $M$ linear inequality constraints, resulting in a total of $2^N \times M$ linear constraints. Because of this unfavorable scaling, this formulation quickly becomes intractable even with specialized linear optimization methods.
	
	We therefore introduce a new reformulation that significantly simplifies the problem by exploiting the fact that, for each hull facet, we can use the facet's normal vector to pre-compute which of the $2^N$ box vertices will be the first to violate the constraint when the box is expanded. This allows us to eliminate the need to check all $2^N$ box vertices for each facet, reducing the number of total linear constraints to simply $M$ and restoring tractability. This reformulation allows us to solve the robust optimization problem in seconds on a laptop by making use of the AeroSandbox optimization framework~\cite{sharpeAcceleratingPracticalEngineering2024} with the interior-point solver~\cite{wachterImplementationInteriorpointFilter2006}.
	
	The objective function $\Phi$ maximizes the volume of the box uncertainty set $\Phi(\mathbf{\Delta x}) = V(\mathcal{B}) = 2^N \prod_{i=1}^N \Delta x_i$. This can be conveniently expressed as a linear function of design variables if the underlying box-side-length variables are log-transformed, corresponding to a cost function that penalizes the relative tolerance of each dimension equally. Without loss of generality, this objective function could be modified to include a linear reweighting of log-transformed dimensions to emphasize that achieving a given relative tolerance level may correspond to different costs for certain dimensions.
	
	The choice to formulate the problem via a convex level set brings significant advantages during the subsequent optimization step, namely that (a) highly efficient convex optimization methods can be used, and (b) any converged solution is provably globally optimal (with respect to the convex-relaxed problem). However, this relaxation process can introduce modeling error if the true performance level set is nonconvex, as illustrated in \cref{fig:optimization}. For design problems with smooth performance functions, a Taylor-series argument can be made to suggest that, even in such cases, this error tends to be small for small box-sets. This seems to be true for the \ac{WG-CPW} transition design described in this work: of the 351 points sampled with the Bayesian \ac{LCB} approach for the feasibility boundary determination problem, 107 of those points are outside $\mathcal{F}$ (i.e. have a performance worse than the threshold), but none of those points outside $\mathcal{F}$ lie inside the convex hull of the 244 sampled points within $\mathcal{F}$. However, this is not guaranteed, so estimating the degree of level-set nonconvexity (if any) may be warranted when applying this procedure to other applications.
	
	The final design  $\left(\mathbf{x^\star}, \mathbf{\Delta x^\star} \right)$ for the transition fabricated in this work was determined assuming a waveguide bulk resistivity of \qty{2.6e-6}{\ohm \centi \meter} and a loss tangent of \num{1.5e-5} in the \ac{CPW} dielectric. We used a loss goal of $L_\text{goal}=\qty{-30}{\dB}$ for the na\"ive optimization and a lower $L_\text{goal}=\qty{-20}{\dB}$ for the tolerancing operation. We tried several values of $f_0$ until we found that an acceptable performance threshold of $f_0=\qty{-12}{\dB}$ gave a tolerance similar to that of our chosen 3D-printing manufacturing process. The optimized parameters for this design $x^\star_i$ are specified in \cref{tab:params} in \cref{app:geometry}; other design parameters were held fixed during optimization. A box $\mathcal{B}$ with identical side-lengths (for simplicity) was used to calculate the required tolerance on $x^\star_i$, which was determined to be
	\begin{equation*}
		\Delta x^\star = \Delta x^\star_i = \text{\qty{7.9}{\micro \meter}}, \qquad i=1,\ldots,N
	\end{equation*}
	where $N = 11$. This tolerance also happens to keep the simulated transmitted power above 95\% over an octave bandwidth for 95\% of randomly sampled transitions within tolerance~\cite{stoverVerticalIntegratedSpectrometer2024, stoverHighdensityVerticalIntegrated2025}.

	\section{Fabrication}
	\label{sec:fab}
	
	\begin{figure}[t]
		\centering
		\includegraphics[width=\linewidth]{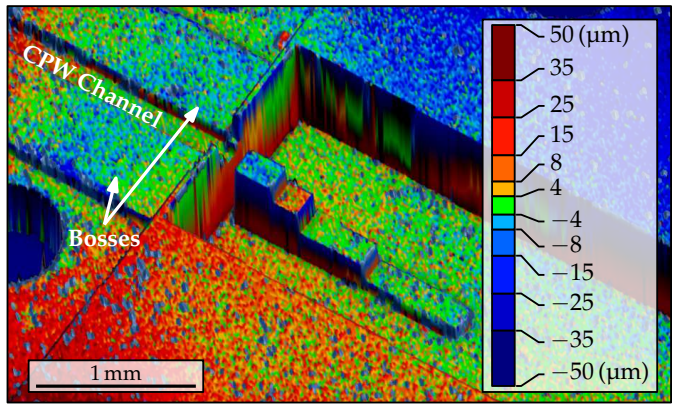} %
		\caption{Confocal micrograph of a typical 3D-printed \acl{WG} transition structure, including surface deviations from the nominal design. Model and measured surface alignment is performed with respect to the step structure. Features with surfaces marked in green ($\pm\text{\qty{4}{\micro\meter}}$) must be within the feature size tolerance of $\pm\Delta x^\star \approx \text{\qty{8}{\micro\meter}}$. However, surface deviations $\pm\text{\qty{8}{\micro\meter}}$ or more may still be within tolerance. Note that surfaces further from the surface alignment features have worse deviations. This illustrates the difficulties of bump bonding across large 3D-printed surfaces.}%
		\label{fig:Confocal}
	\end{figure}
	
	\begin{figure}[t]
		\centering
		\includegraphics[width=\linewidth]{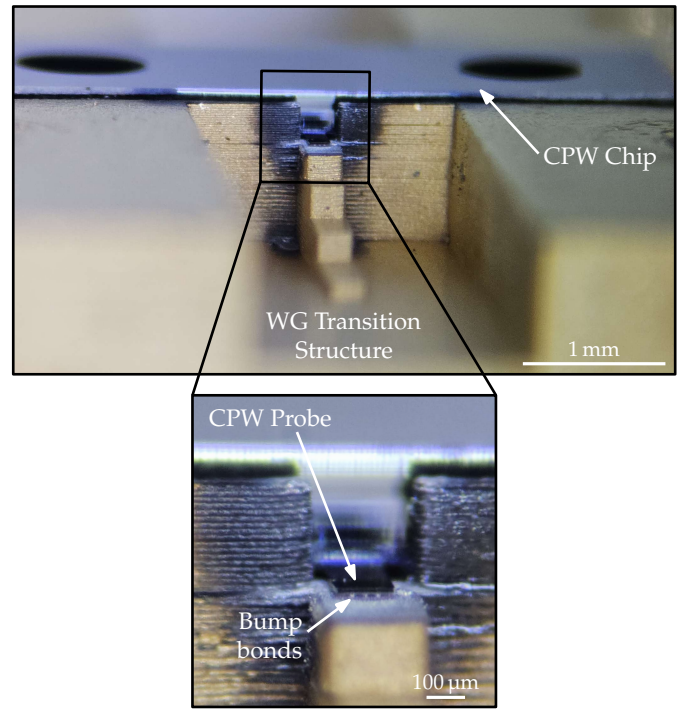}
		\caption{Photograph of a completed \acl{WG-CPW} transition with detail of the \acl{CPW} probe. The bump bonds which electrically connect the \acl{CPW} conductor trace to the top step of the \acl{WG} transition step structure are visible in the detail. The discoloration around the \acl{CPW} probe on the \acl{WG} transition structure was caused by heating from thermal deposition of the indium. It does not impact the resistivity.}
		\label{fig:WG-CPW_Transformer}
	\end{figure}
	
	Two major components comprise this \ac{WG-CPW} transition: A three-dimensional waveguide transition structure and a die containing the \ac{CPW} with extruded cantilevers electrically bonded to the waveguide transition structure. The electrical interfaces between these components are made with indium bump bonds. Indium is deposited both as pads on the waveguide transition and bumps on the die, with the thickness of this deposition chosen to enable interconnection while accounting for the tolerance of the waveguide structure, a compression of $\unsim50\%$ during bonding, and tolerance of misalignment. 
	
	\subsection*{Waveguide Transition Structure}
	The waveguide transition structure was fabricated by a commercial vendor\footnote{BMF Precision, Inc., (Maynard, MA, USA).} through additive manufacturing with alumina ceramic and then made conductive through a metallization process. Photographs of this structure are shown in \cref{fig:WG-CPW_Transformer}.

	In detail, the waveguide structure was fabricated by projection micro-stereolithography~\cite{geProjectionMicroStereolithography2020}. A photosensitive resin containing a slurry of \qty{500}{\nano \meter} alumina particles was selectively hardened layer-by-layer by ultraviolet light, then subjected to a debinding and sintering process that leaves behind only alumina ceramic. 
	
	This alumina substrate was then metallized following a procedure developed at NIST for silicon platelet feed-horn arrays~\cite{nibarger84PixelAllSilicon2012}. The process starts with an in-situ sputter etch followed by DC magnetron sputtering a \qty{200}{\nano \meter} thin film of titanium to promote adhesion; this process was performed by a commercial vendor.\footnote{LGA Thin Films, Inc., (Santa Clara, CA, USA).} Then a \qty{1}{\micro \meter} copper layer is sputtered to make the surface conductive, and a \qty{5}{\micro \meter} gold layer is electroplated to realize a high-conductivity surface impervious to oxidation--this was performed by a separate vendor.\footnote{Custom Microwave, Inc., Vitesse Systems (Longmont, CO, USA).}
	
	To perform bump bonding, indium pads are thermally evaporated onto the \ac{WG} transition. The target thickness of these pads is \qty{10}{\micro \meter} to provide sufficient allowance for bonding and compression; they were measured to be \qtyrange{9}{15}{\micro\meter} thick by confocal microscopy. The indium is deposited in three regions where the bump bond connections are made. First, it is deposited onto the top step of the stepped impedance transformer structure to make the \ac{CPW} conductor trace connection. It is also deposited onto the lower surface of the rectangular aperture in the backshort, on either side of the \ac{CPW} trough, to make the \ac{CPW} ground plane connection, as well as onto the bosses straddling the length of the \ac{CPW} trough, for bonding with the \ac{CPW} ground traces to create the air-bridge that connects the \ac{CPW} grounds. The indium deposition is controlled with a 3D-printed titanium hard mask that fits over the part for the deposition process. This mask is fabricated from titanium using direct metal laser sintering, making it thermally and mechanically robust and reusable. The hard mask includes a drafted sleeve that locates the waveguide component to within $\pm$\qty{50}{\micro\meter} in either direction beneath the mask apertures when the component is placed into the mask. This tolerance is sufficient because on the waveguide transition structure, only pads must be deposited, not individual bumps.
	
	For evaporation of indium during thermal deposition, a low-mass tungsten boat was used to hold the indium. Ceramic crucibles, which have a lower thermal conductivity, were found to radiate too much heat onto the sample, since they require a higher temperature from the heating source to achieve the same evaporation rate as a tungsten boat. Multiple depositions alternated with cooling periods were performed to keep temperatures lower and refill the crucible in order to achieve the desired indium thickness. The sample holder was also actively water-cooled. Still, the surface of the \ac{WG} transition heated significantly---this is evident from the discoloration around the deposition area~(\cref{fig:WG-CPW_Transformer})---since the alumina body of the transition is an insulator, and the thin metallization does not conduct heat substantially. The discoloration appears to be superficial; it does not impact the surface resistivity. Use of a semiconductor device analyzer and probe station showed that both pristine and discolored surfaces have ohmic I-V curves with the same resistances at the same probe spacings.
	
	\subsection*{Die Fabrication}
	The die carries a \ac{CPW} with conductors made of a gold thin-film on the \qty{15}{\micro \meter} thick device layer of a diced \ac{SOI} wafer. Bump bonds are deposited onto the gold to facilitate electrical interconnections to the waveguide transition structure. The input and output interconnects lie along two opposing \qty{150}{\micro \meter} wide by \qty{401}{\micro \meter} long and \qty{15}{\micro \meter} thick cantilevers. The cantilever is made by first etching away the device layer around the cantilever planform before flipping the wafer around and etching away the handle and oxide layers of the \ac{SOI} underneath and around the cantilever. The die and bump bonds are shown in \cref{fig:InAndSWIR}. We now describe the process used to fabricate this structure. 
	
	\begin{figure}[t]
		\centering
		\includegraphics[width=\linewidth]{Bump_bonding_4.tikz}
		\caption{Optical micrograph of \acl{CPW} chip (top right) as viewed from the device side, with details of \ac{SEM} images of the indium bumps (top left, bottom left) and a composite image of the die during alignment as viewed from the handle side through the \acl{SWIR} microscope during bonding (bottom right). The transparency of crystalline silicon to infrared wavelengths allows the traces deposited on the device side to be seen from the handle side, enabling alignment between device-side alignment marks and physical features on the \acl{WG} transition structure below.}
		\label{fig:InAndSWIR}
	\end{figure}

	The fabrication of the \ac{CPW} chip entailed the following steps:
	\begin{enumerate}
		\item A lift-off process was used to pattern the \ac{CPW} conductor and ground traces onto the device side of the \ac{SOI} wafer. The \ac{SOI} comprises a \qty{15}{um} device layer, \qty{0.5}{um} buried oxide layer, and \qty{300}{um} handle layer. Before each lithography step, the wafer is cleaned using sonication in acetone and isopropyl alcohol baths, a \ac{DI} water rinse, and a spin dry. A negative photoresist is spun onto the wafer, followed by exposure with a maskless aligner, development, and a downstream oxygen descum. A \qty{15}{nm} titanium seed layer and \qty{200}{nm} gold layer are then deposited with electron-beam physical vapor deposition. Afterwards, the wafer is left in \ac{NMP} at \qty{80}{\degreeCelsius} overnight to complete the lift-off. %
		
		\item The alignment markers are then transferred from the device layer to the handle layer for the later thru-handle etch. A quick (\qty{5}{\cycle}) \ac{DSE} into the back of the wafer is performed using a positive photoresist that is patterned with the maskless aligner using backside alignment.
		
		\item Indium is deposited with thermal evaporation into bumps over the \ac{CPW} traces using the same evaporation process as that of the indium deposition on the waveguide transition structure. A \qty{20}{\micro\meter} AZ P4620 positive photoresist\footnote{AZ P4620 Positive Thick Photoresist (Merck KGaA, Darmstadt, Germany).} mask is created using a double-spin process\footnotemark{} to form square bump bonds \qtyproduct[product-units = single]{21.5x21.5}{\micro\meter} in area and with a target thickness of \qty{10}{\micro\meter} (and measured by confocal microscopy to be \qtyrange{10}{13}{\micro\meter} tall). Again, thermal deposition is performed with active water cooling. The higher thermal conductivity of the silicon \ac{SOI} (as compared to the alumina waveguide structure) allows its surface temperature to be kept well below the melting point of the photoresist. Lift-off is performed with the deposited indium facing downwards for a good release; the wafer is clamped to a teflon tripod to keep it from touching the bottom of the beaker. When lift-off was attempted with the deposited indium facing up, the indium redeposited onto the wafer surface, adhering to it and preventing successful lift-off.
		
		\footnotetext{The double-spin process is as follows: 
			\begin{enumerate*}[
				label=(\roman*),
				itemjoin={{; }},
				itemjoin*={{; and }}
				]
				\item Spin at \qty{2500}{\rpm} at \qty{500}{\rpm} for \qty{45}{\second}
				\item Bake at \qty{110}{\degreeCelsius} on a hot plate for \qty{2}{\minute}
				\item Let cool \qty{30}{\second}
				\item Repeat the spin.
				\item Bake at \qty{115}{\degreeCelsius} on a hot plate for \qty{3}{\minute}
				\item \qty{20}{\second} at \qty{100}{\thou}
				\item Let rehydrate for \qty{30}{\minute} before exposure
			\end{enumerate*}.}
		
		\item A \ac{DSE} is then performed on the device side of the \ac{SOI} wafer to form the \ac{CPW} probe, using the same \qty{20}{\micro\meter} AZ P4620 positive photoresist double-spin process for the mask. A slower spin acceleration of \qty{50}{\rpm \per \second} is used to ensure the integrity of the bump bonds.
		
		\item The device side of the wafer is adhered onto a carrier wafer in order to etch the handle side, in a process informed by the work of Tang et al.~\cite{tangFabricationOMTCoupledKinetic2020}. A solution of ethylene glycol-phthalic anhydride resin (``mounting wax'')\footnote{PELCO Quickstick 135 Temporary Mounting Wax, (Ted Pella, Inc., Redding, CA, USA).} and \ac{PGMEA} with a 1.5:1 mass ratio is prepared to form the adhesive. To dissolve the resin, it is cut into pieces approximately \qty{3}{\milli \meter} in size before being submerged in \dac{PGMEA} and sonicated at \qty{40}{\kilo \Hz} on medium power in an \qty{80}{\degreeCelsius} bath for about \qty{1}{\hour}. A carrier wafer is then roughened by a fluorine \ac{RIE} to enhance adhesion; without this step, the resin coating beaded up on the silicon surface. The \ac{RIE} is performed with a high bias power of \qty{500}{\watt} and a lower \ac{RF} power of \qty{100}{\watt} in order to physically sputter the surface. The resin solution is then spun onto the carrier wafer at \qty{800}{\rpm}, which is placed into a vacuum oven---this step is repeated until a resin thickness of \qty{37}{\micro\meter} is achieved (enough to account for the device thickness and bump bond heights), which requires three spins and vacuum bakes. The device side of the SOI is then aligned with the carrier wafer and pressed into the wax with a \qty{1.87}{\kilo \gram} weight before being placed in an \qty{11}{\torr} vacuum at \qty{110}{\degreeCelsius} for \qty{30}{\minute}. 
		
		\item A backside \ac{DSE} is performed, this time on the handle side of the \ac{SOI} wafer, to undercut the \ac{CPW} probe and form the bulk of the \ac{CPW} die. A \qty{27}{\micro\meter} thick AZ 40XT positive photoresist\footnote{AZ 40XT-11D Chemically Amplified Thick Photoresist (Merck KGaA, Darmstadt, Germany).} mask is used, with an \qty{85}{\degreeCelsius} post-development hard bake\footnotemark{} to increase the selectivity of the \ac{DSE}. A low heat exchanger temperature of \qty{12}{\degreeCelsius} is used to ensure high aspect ratio and selectivity, and every \qty{300}{\cycle}, a \qty{15}{\minute} wait period is included to allow the wafer to cool back down, again ensuring high selectivity even with the poor thermal conduction through the mounting wax. %
		
		\footnotetext{The hard bake is performed with programmable lift pins and the following lengths of time at each pin height:
			\begin{enumerate*}[
				label=(\roman*),
				itemjoin={{; }},
				itemjoin*={{; and }}
				]
				\item \qty{30}{\second} at \qty{500}{\thou}
				\item \qty{20}{\second} at \qty{100}{\thou}
				\item \qty{20}{\second} at \qty{25}{\thou}
				\item \qty{300}{\second} directly on the hot plate, with a vacuum
				\item \qty{200}{\second} at \qty{25}{\thou}
				\item \qty{20}{\second} at \qty{100}{\thou}
				\item \qty{20}{\second} at \qty{100}{\thou}
			\end{enumerate*}.}
		
		\item Finally, the oxide layer exposed by the backside etch is removed by \qty{25}{\minute} of moderate agitation in buffered oxide etch. After dicing, the dies are released from the carrier wafer by sonication in acetone on medium-low power at \qty{72}{\kilo \Hz} for \qty{4}{\minute}, which dissolves the mounting wax. Sonication at a lower \qty{40}{\kilo \Hz} frequency was attempted, but during this step, even at the lowest power available, the \ac{CPW} probe cantilevers consistently broke off the die---the higher frequencies did not damage the cantilevers. The dies are oriented vertically during this step to allow the carrier layer to slide off without damaging the cantilevers, and they are removed from the acetone \qty{30}{\second} afterwards in order to fully dissolve any residual mounting wax.
	\end{enumerate}
	
	\section{Assembly}
	\label{sec:bond}
	
	The waveguide transition structure and die are assembled using a combination of flip-chip bonding for electrical interconnections and epoxy for mechanical strength. The transition is completed by mounting it into a custom-machined aluminum part that includes a standard waveguide flange, in order to provide an interface to test equipment, as well as a metal cap which is used to seal the top of the waveguide~(\cref{fig:DUT}).
	
	\subsection*{Bump Bonding}
	
	\begin{figure}[t]
		\centering
		\includegraphics[width=\linewidth]{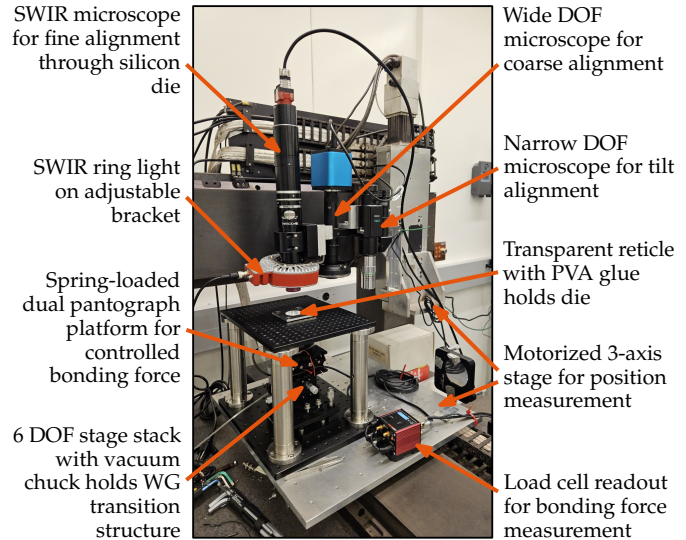}
		\caption{Photograph of the custom flip-chip bonder. Conventional flip-chip bonders often use autocollimators to align a device chip to a flat, optically smooth substrate; this approach is infeasible for bonding a chip to a 3D-printed substrate. Instead, a narrow \ac{DOF} microscope is used to measure the vertical distance between the chip and substrate with a motorized 3-axis stage for pitch and tilt alignment, and a \ac{SWIR} microscope is used to see through the silicon chip to alignment markers on the other side to perform the final translational and rotational alignment.}
		\label{fig:bonder}
	\end{figure}
	
	Conventional flip-chip bonders often use autocollimators to ensure die coplanarity between the two substrates; however, these instruments require both dies to have optically flat surfaces with larger areas than is available with the waveguide transition structure. Ensuring coplanarity between the relatively small and textured surface of the 3D-printed waveguide transition substrate and the die requires a different approach.
	
	To this end, we built a custom flip-chip bonder (\cref{fig:bonder}). This bonder includes a motorized 3-axis stage which holds 3 microscopes: a wide depth-of-field telecentric microscope for overall alignment, a shallow depth-of-field microscope to perform precision vertical measurements for pitch and tilt alignment, and a \ac{SWIR} microscope to permit fine translational and rotational alignment. Since monocrystalline silicon is transparent to \ac{SWIR} light, such a microscope can view alignment markings on the underside of a silicon chip from the top and even view through the silicon to detect reflective features underneath, which makes fine alignments with alignment markings on the far surface of the silicon chip possible.
	
	The flip-chip bonder also has a six \ac{DOF} stage stack controlled by Vernier micrometers: a lower stage controls rotation, pitch, and roll, and an upper stage controls translation. A spring-loaded dual-pantograph platform sits on top of the upper stage. Its purpose it to provide linear deformation along the bonding axis while remaining stiff in other \ac{DOF}s in order to provide a controlled force during bonding. The platform was modified from a dual pantograph "lab jack" to make it free running, with a compression spring installed between adjustable set screws on the top and bottom platform, to pre-load the platform as necessary, and a 3D-printed sleeve to guide the spring along part of its length. A load cell fastened to the top of this platform is used for bond force readout. A vacuum chuck affixed to the load cell holds the 3D-printed waveguide substrate during bonding.
	
	A platform with a large central aperture is rigidly held above the stage stack. A removable custom machined chuck with a smaller aperture fits into the center of the platform and holds a glass slide or reticle beneath it mechanically with specimen clips. The die is affixed to the slide using water soluble \ac{PVA} glue\footnote{Elmer's Washable Liquid School Glue (Elmer's Products, Inc., Westerville, OH, USA)} in a solution of \ac{DI} water with a 1:1 mass ratio, in order to achieve the desired viscosity. Higher viscosities were found to cause the die to set at an angle unsuitable for bonding.
	
	We achieve pitch/tilt alignment by using the motorized 3-axis stage to measure the vertical positions of the substrate and die at which they come into focus under the shallow depth of field ($\pm\text{\qty{1.6}{\micro \meter}}$) microscope objective. The substrate and die are aligned when they are found to be equidistant at several distant points on their surfaces. The lower stage in the stage stack controls rotation, pitch, and roll via manual micrometer adjustment. A semi-automated program was developed that assists in performing the pitch-tilt measurement--it moves the motorized stage automatically to the next measurement location once a measurement has been made, and calculates the pitch and roll micrometer adjustments needed to parallelize the substrate and die. The program also computes variances given the microscope depth of field or multiple measurements to further decrease error, and propagates these variances into an estimation of the final misalignment, with typical misalignment variances approaching $\pm\text{\qty{3}{\micro \meter}}$. We found, however, that systematics such as warped surfaces can cause substantial additional misalignment and must be accounted for, which we discuss in \cref{sec:meas}.
	
	Coarse translational and rotational alignment is performed by aligning parallel features and measuring distances between features on the substrate and those on the die, while adjusting the rotation of the lower stage and position of the upper stage. The final alignment is performed with the \ac{SWIR} microscope, by aligning edge features on the substrate with imaging alignment marks lithographically deposited onto the underside of the die.
	
	Just before bonding, an optically transparent epoxy is applied to the substrate to ensure mechanical robustness of the bond to handling. The bonding is performed by raising the lower stage until a bonding force of $\unsim\qty{7}{kgf \per mm^2}$ of bond area is achieved. After the epoxy cures, the vacuum is deactivated and the adjustable stage is lowered. The slide, substrate, and die are then submerged in \ac{DI} water with gentle agitation until the die and substrate separate from the slide and the residual \ac{PVA} glue is fully dissolved.

	\subsection*{Assembly of the Test Device}
	The test device is completed by closing each waveguide with a planar metal cap. The cap is custom-machined out of aluminum and extrudes from a UG-387/U waveguide flange with a rectangular aperture about the width and height of the transition structure in the center. The waveguide end of the transition structure fits into this aperture, which also locates it. Each cap is mechanically affixed to the test device using a pair of bolts that passes through the cap and the \ac{WG} test structure. During assembly the test device is aligned so that the flange face is coplanar to the face at the waveguide end of the transition structure. The final step is to mount the \ac{WG} flanges onto the assembly. The completed test structure is shown in \cref{fig:DUT}.
	
	\begin{figure}[t]
		\centering
		\includegraphics[width=\linewidth]{DUT_P1_S11.tikz}
		\caption{Image of the measurement setup for the upper band (\qtyrange{110}{170}{\GHz}), including the \ac{VNA} mixer head, waveguide transitions, and \ac{DUT}, which comprises the waveguide mounting flange with a waveguide cap, and \ac{WG-CPW} transition. The lower band (\qtyrange{75}{110}{\GHz}) was measured with a similar setup. Due to the high losses through the transition, the termination boundary condition is unimportant; the transition was characterized with reflectivity measurements and the loss model described in \cref{sec:meas}.}
		\label{fig:DUT}
	\end{figure}
	
	\begin{figure*}[t]
		\centering
		\input{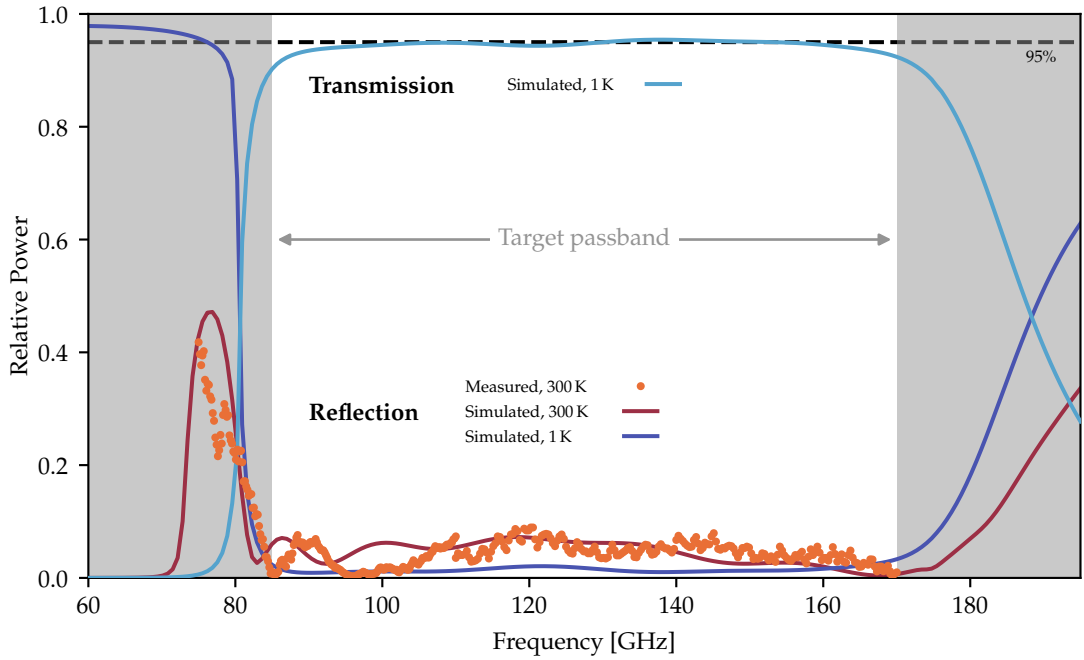}
		\caption{Comparison between measured and simulated reflected power for the transition at room temperature. Simulations of the reflected and transmitted power at cryogenic temperatures are included. The measurements from \qtyrange{75}{110}{\GHz} and \qtyrange{110}{170}{\GHz} were taken with two separate \acp{VNA}.}
		\label{fig:Reflectivity}
	\end{figure*}
	
	\section{Measurements}
	\label{sec:meas}
	
	A \ac{VNA} was used to characterize the room-temperature performance of a transition using a Short-Open-Load calibration. A different \ac{VNA} was used for the lower (\qtyrange{75}{110}{\GHz}) and upper (\qtyrange{110}{170}{\GHz}) frequency ranges, which appears in the measurements as a difference in relative noise amplitude (orange dots in \cref{fig:Reflectivity}). We note that the converter for the upper frequency range was damaged, which explains its anomalously high noise. The resulting data are shown in \cref{fig:Reflectivity}. These data show less than 10\% of reflected power across the full octave target band at room temperature.
	
	These data are compared to simulations of the reflected power at \qty{300}{\kelvin} and simulations of the reflected and transmitted power at \qty{1}{\kelvin}. The room temperature simulated reflection shows reasonable agreement at the 5\% level with warm reflection measurements. The inferred transmission indicates around 95\% coupling efficiency at cryogenic temperatures and a half power bandwidth ratio of 2.3. Such a bandwidth is comparable to that of the waveguide-to-microstrip \acp{OMT} on the pixels of \dac{ACT}, \ac{SO}, and other experiments~\cite{mcmahonMultichroicFeedHornCoupled2012}.
	
	The waveguide transition that was measured has two defects. First, it has a chipped region in the top left corner of the aperture in the backshort wall, as viewed from the waveguide opening, of approximately triangular shape and \qty{377}{\micro \meter} in length by \qty{234}{\micro \meter} in height (see \cref{fig:measured_device} in \cref{app:measured}). This chip was the result of a poor release from the indium deposition hard mask due to insufficient allowance in the hard mask design---future iterations of the hard mask increase this allowance. The chipped area of the backshort widens the aperture; although the dimensions of the new widened backshort aperture still have a waveguide cutoff higher than the highest frequencies in the target passband, some radiative leakage due to tunneling still occurs. Simulations show that this changes the reflected power over the pass-band by up to 2.5\%, thus we include the chipped backshort in the warm simulations shown in  \cref{fig:Reflectivity}. Since we expect to cure this defect in future iterations, we choose to exclude it from the cold simulations.
	
	Second, the \ac{CPW} probe on the measured transition has a lateral offset of \qty{26}{\micro \meter} to the left when viewed from the waveguide opening, such that only one of the \ac{CPW} ground traces connects to the backshort. This offset was caused by imperfect pitch alignment of the die around the transmission line axis greater than the height of the bump bonds, causing a lateral shift when pressure was applied during bonding. This offset could be alleviated by a change in the pitch alignment methodology. Before bonding, substrate heights are measured beyond the lateral edges of the die (since the die blocks the substrate underneath it), then the die heights are measured, and the differences compared to perform the pitch alignment. However, the surface of the 3D-printed substrate is affected by warping due to the sintering process, which causes an apparent vertical deviation the further away from the transmission line axis which can cause misalignment. Instead, we can use the translation stage to move the die away from the desired measurement location before a substrate height measurement, and then translate it back, in order to perform distance measurements between adjacent locations on the substrate and the die. This entails making measurements on the substrate bosses, where warping remains negligible, as seen in \cref{fig:Confocal}. This solution was only determined after the completion of the measurement, since the defect only occurred intermittently during development.
	
	Simulations show that this lateral offset causes deviations in the reflected power below the no-offset case by \qtyrange{1}{3}{\decibel} above \qty{140}{\GHz} in the lossy, room-temperature case. This misalignment was therefore included in calculating the simulated room temperature reflection. Future cryogenic transitions must ensure that the indium bonds on both \ac{CPW} grounds connect to the backshort, since simulations show that having both interconnections is required to maintain the 95\% transmission when other losses are low. This imposes a constraint on the lateral offset of less than \qty{18}{\micro \meter}. However, if both \ac{CPW} grounds connect to the backshort, simulations show negligible performance improvement with smaller lateral offsets.
	
	Several bonds were made without these defects, such as that shown in \cref{fig:WG-CPW_Transformer}, but these bonds were made with dies that did not have the high quality \ac{CPW} conductor trace needed for measurement. 
	
	\subsection*{Simulation and Characterization}
	\label{sec:sim}
	
	The simulations in \cref{fig:Reflectivity} include conductive and dielectric losses and account for the roughness of the metallized gold waveguide as measured via confocal microscopy. 
	
	The dielectric loss was estimated by a Drude model~\cite{ashcroftSolidState1976}, which has been shown to be valid for n-type silicon with the measured DC resistivity around \qty{100}{\GHz}~\cite{kinasewitzInvestigationComplexPermittivity1983}. The silicon loss tangent is calculated as
	\begin{equation}
		\tan{\delta} \equiv \frac{\epsilon''}{\epsilon'} \\
		\label{eqn:losstan}
	\end{equation}
	where
	\begin{equation*}
		\begin{aligned}
			\epsilon &= \epsilon' + i\epsilon'' = \epsilon_0\epsilon_r + i\epsilon'' \\
			\epsilon'' &= \frac{1/\rho_0}{\omega(1+\omega^2\tau^2)}
		\end{aligned}
	\end{equation*}
	and the DC resistivity $\rho_0$ of the silicon was measured with a 4-point probe on another wafer in the same batch and determined to be \qty{4.0 \pm 0.1}{\ohm \cm}. The relaxation time $\tau$ was calculated as
	\begin{equation*}
		\tau = \frac{m_c}{\rho_0 n e^2} = \qty{6.8e-14}{\second}
	\end{equation*}
	where $m_c = 0.26m_e$ is the conductivity effective mass~\cite{kinasewitzInvestigationComplexPermittivity1983}, $e$ is electron charge, and $n$ is the carrier density, which is determined from the DC resistivity~\cite{astminternationalConversionBetweenResistivity1988}. Since $\omega^2 \tau^2 \unsim 10^{-4}$, the relaxation time is negligible, so the imaginary component of permittivity used for the dielectric model is
	\begin{equation}
		\epsilon'' \approx \frac{1}{\rho_0 \omega} \\
		\label{eqn:losstaneps}
	\end{equation}
	which gives a value of $\tan{\delta} = \num{0.30}$ at \qty{127.5}{\GHz}. When the loss tangent was varied, this loss tangent value also provided the closest qualitative agreement between the measured and simulated reflectivity by visual inspection.
	
	The losses in the 3D-printed and metallized waveguide transition due to the surface roughness (caused by the 3D-printing and metallization process) and the finite conductivity of the metallized gold were estimated by a causal Huray loss model~\cite{brackenCausalHurayModel2012}, which models the surface as a conglomeration of spheres. This choice of roughness model was motivated by the large scale of the surface roughness of the transition ($S_a=\text{\qty{0.5}{\micro\meter}}$ and $S_q=\text{\qty{0.6}{\micro\meter}}$, similar to the alumina particle size) relative to its skin depth $\delta_{sd} = \sqrt{2/\omega \sigma \mu} \approx \text{\qty{0.2}{\micro\meter}}$. The choice of the Huray model was also motivated by confocal microscopy measurements of the surface of the transition, which show a conglomeration of spherical nodules (see \cref{fig:surface_sample} in \cref{app:surface}). The model coefficients---nodule radius and number of nodules per unit area---were determined from these measurements of a typical patch of this surface (\cref{app:surface}). Since the nodule radius was found to vary from \qtyrange{5}{15}{\micro \meter} but the model implemented uses only one nodule size, simulations were performed with the nodule radius swept across the measured range and the best fit was chosen. This choice primarily affects the reflectivity at the waveguide low frequency cutoff of \qty{80}{\GHz} due to the pole in the attenuation of rectangular waveguides as a function of the wavenumber near the cutoff~\cite{pozarMicrowaveEngineering2012}, so the reflectivity below \qty{85}{\GHz} was chosen as the fit criterion. The resulting simulated reflectivity, using the \qty{7}{\micro \meter} nodule radius and \qty{4.0}{\ohm \cm} dielectric resistivity, is compared to the measured reflectivity in \cref{fig:Reflectivity}.
	
	This best-fit material model of the electroplated gold is then used to estimate the transmission of the transition at cryogenic temperatures (\qtyrange{1}{4}{\kelvin}). The total resistivity of gold at \qtyrange{1}{4}{\kelvin} as measured by~\cite{matulaElectricalResistivityCopper1979} is multiplied by a correction factor of 3 for electroplating impurities, as measured by~\cite{bernatThermalElectricalConductivities2007}, to get an estimated bulk resistivity of \qty{6.6e-8}{\ohm \cm}. This resistivity is then applied to the same causal Huray model to simulate the transition performance as a function of temperature. These simulations imply a 95\% transmitted power in the \qtyrange{90}{165}{\GHz} band at cryogenic temperatures.
	
	\begin{table}[t]
		\caption{\Acl{WG-CPW} transition parameter values. The parameters $x^\text{min}_i$ give the solution to na\"ive optimization with $\argmin{f(\mathbf{x})}$, and the parameters $x^\star_i$ give the solution to the tolerance-aware optimization problem described by \cref{eqn:optalg}. These optimized parameters are given in the left table. The other values were held constant during this optimization (right table). $W$ is the \acs{CPW} conductor trace width, $G$ is the distance from the ground plane to the \acs{CPW} conductor trace, and $H$ is the device layer thickness. All parameters listed below correspond to the parameters of the same name in the waveguide to microstrip transition (\cref{fig:uStrip_Diagram}).}
		\begin{tabular}[t]{@{} l S[table-format=3.1] S[table-format=3.1] @{}}
			\toprule
			\thead{Param.} & \multicolumn{1}{c}{\thead{$x^\text{min}_i$\,{}[\unit{\micro\meter}]}} & \multicolumn{1}{c}{\thead{$x^\star_i$\,{}[\unit{\micro\meter}]}} \\
			\midrule
			$x_1$    & 370 & 362      \\ %
			$x_2$    & 243 & 248       \\ %
			$x_3$    & 591 & 590      \\ %
			$x_4$    & 871 & 875      \\ %
			$x_5$    & 206 & 190      \\ %
			$x_6$    & 216 & 220      \\ %
			$x_7$    & 211 & 191      \\ %
			$x_8$    & 84  & 97       \\ %
			$x_9$    & 100 & 98       \\ %
			$x_{10}$ & 112 & 102       \\ 
			$x_{11}$ & 33 & 50        \\ 
			$\Delta x$& \multicolumn{1}{c}{---} & 7.9 \\
			\bottomrule
		\end{tabular}
		\hfill
		\begin{tabular}[t]{@{} l S[table-format=3.1] @{}}
			\toprule
			\thead{Param.} & \multicolumn{1}{c}{\thead{Value\,{}[\unit{\micro\meter}]}} \\
			\midrule
			$a$   & 1879\\ %
			$b$   & 964 \\ %
			$t$   & 272 \\ %
			$w_p$ & 150 \\ 
			$l_m$ & 401 \\ 
			$t_b$ & 150 \\ %
			$w_a$ & 250 \\ 
			$w_t$ & 100 \\ 
			$h_t$ & 100 \\ 
			$W$   & 2.0 \\ %
			$G$   & 5.3 \\ %
			$H$   & 15.0\\
			\bottomrule
		\end{tabular}
		\label{tab:params}
	\end{table}
	
	\section{Conclusion}
	\label{sec:conclusion}
	
	\begin{figure}[t]
		\centering
		\includegraphics[width=\linewidth]{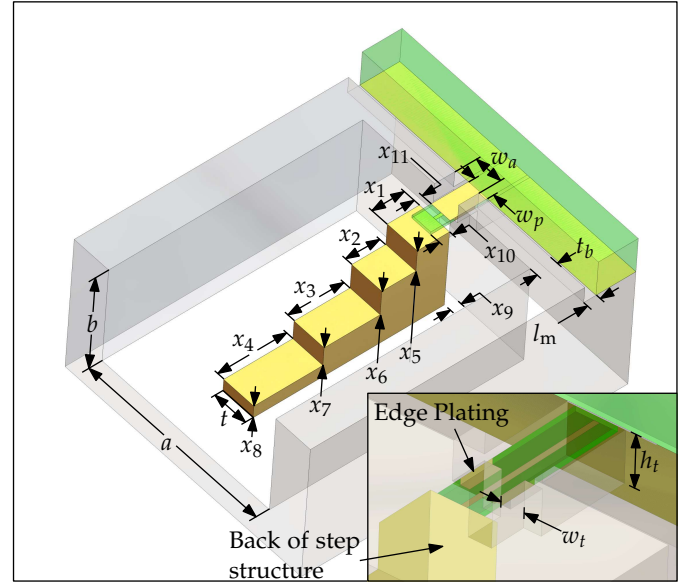}
		\caption{Diagram of the \acl{WG} to microstrip transition design, with detail of the microstrip port, as seen from the back side, inset. The \acl{WG} cap is omitted. Translucent green denotes silicon, yellow represents traces or metallized surface, and, apart from the step structure, the metallized waveguide structure is transparent and colorless so that underlying details may be seen. Dimensions are given in \cref{tab:ustrip_params}. All dimensional parameters shown in this figure also apply to the \ac{WG-CPW} transition.}
		\label{fig:uStrip_Diagram}
	\end{figure}
	
	\begin{figure}[t]
		\centering
		\input{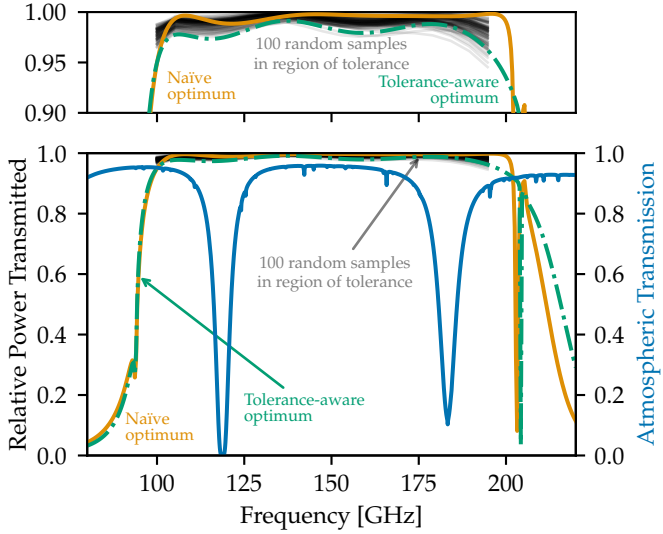}
		\caption{Plot of the relative transmitted power $|S_{21}(\nu)|^2$ for a \acl{WG} to microstrip transition designed for a \qtyrange{100}{200}{\GHz} passband with lossless conductors and the same low-loss dielectric material as assumed in \cref{sec:opt_implementation}. The na\"ive optimum (orange, dash-dotted) was evaluated at $\argmin{f(\mathbf{x})}$ whereas the tolerance-aware optimum was evaluated at $\mathbf{x}^\star$ (green, solid). One hundred random samples within the uncertainty box-set $\mathcal{B}(\mathbf{x}^\star,\Delta x)$ simulated within the passband are shown in gray. The parameters associated with the optima, $x^\text{min}_i$, $x^\star_i$, and $\Delta x$, are tabulated in \cref{tab:ustrip_params}. \ac{SPT} measurements of atmospheric opacity in the Summer at a \qty{45}{\degree} angle are overlaid in blue for reference, courtesy of J. Z.}
		\label{fig:uStrip_Sims}
	\end{figure}
	
	We have presented a millimeter-wave in-line octave-bandwidth \ac{WG-CPW} transition that enables the dense packing of on-chip spectrometers into a telescope focal plane. Room-temperature testing shows less than 10\% reflected power over an octave of bandwidth. These results show reasonable agreement with simulations based on the measured material properties. This model predicts a transmitted power (coupling efficiency) of around 95\% from \qtyrange{90}{165}{\GHz} and above 90\% across the full \qtyrange{85}{170}{\GHz} passband.
	
	We have developed a novel optimization approach which makes the design more robust to manufacturing tolerances by ensuring that tolerance specifications fall within regions of the design space that meet required performance thresholds. This approach provides a generalizable framework for instrument design problems where the performance loss due to manufacturing deviations from the nominal design is non-negligible.
	
	We also demonstrated a new approach to creating interconnects between additively manufactured structures and lithographically fabricated dies through bump bonding. We anticipate that this approach will have a number of uses beyond the present work.
	
	In future work we will adapt the \ac{WG-CPW} transition design to a low-loss, cryogenic environment and integrate it with on-chip spectrometers in order to realize a high-density mm-wave \ac{IFU} that can achieve sensitivity improvements of one or more orders of magnitude over existing approaches. Such an instrument could enable ambitious mm-wave \ac{LIM} surveys.
	
	\appendix
	\section*{APPENDICES}
	
	\section{Transition Geometry} \label{app:geometry}
	
	Parameter values for the waveguide transition test device are tabulated in \cref{tab:params}.
	
	\section{\ac{WG}-to-Microstrip Transition Design}
	\label{app:microstrip}
	
	The optimization and tolerancing methodology in \cref{sec:opt} can also be applied to the design of a waveguide-to-microstrip transition with a higher frequency \qtyrange{100}{200}{\GHz} passband. The microstrip transition inherits many features from its \ac{CPW} transition sibling, including a ground plane that extends beyond the backshort, a die fabricated from two layers of silicon with a cantilevered probe, a stepped ridge impedance transformer, and a bump-bonded interconnection.
	
	Unlike the \ac{CPW} design, the microstrip ground plane lies on a different plane than the conductor trace. In order to short this ground plane to the backshort, a kind of via is needed that connects the ground plane through the microstrip probe to an exposed ground trace coplanar with the conductor trace that can be bump bonded. While the cantilevered \ac{CPW} probe does not have the width or strength to support through-hole vias, the angled deposition of conductive metal onto the sides of this cantilever can create edge plating that connects the ground trace to the ground plane without the need for through holes~(\cref{fig:uStrip_Diagram}). Additionally, the microstrip design requires a ground plane under the device layer, so \ac{SOI} wafers cannot be used. Instead, a \ac{SOI} flip, bond, and etch process~\cite{hessLowLossMicrostripTransmission2019} can achieve a monocrystalline silicon device layer separated from a handle layer by a ground plane.
	
	This waveguide to microstrip transition was simulated for optimization with lossless conductors and the same low-loss silicon dielectric as was assumed in \cref{sec:opt_implementation}. As in \cref{sec:opt}, Bayesian optimization was used to find the na\"ive optimum $\mathbf{x}^\text{min}$. The loss goal was set to $L_\text{goal} = \qty{-25}{\dB}$ for all steps, and the acceptable loss was set to $f_0=\qty{-12}{\dB}$, which gave the optima plotted in green (dot-dashed line) and orange (solid line) in \cref{fig:uStrip_Sims} configured with the parameters tabulated in \cref{tab:ustrip_params}. For these loss goals and acceptable losses, the maximum allowable tolerance was found to be $\Delta x = \text{\qty{6.2}{\micro \meter}}$. To illustrate the robustness of the tolerance-aware optimum to manufacturing deviations within this tolerance, the tolerance space $\mathcal{B}(\mathbf{x}^\star,\Delta x)$ was randomly sampled; these one hundred random samples are also plotted over the \qtyrange{100}{200}{\GHz} passband in \cref{fig:uStrip_Sims} as gray (translucent black) lines. Without considering losses due to surface roughness or resistivity, the transition achieves around 97\% to 99\% power transmission over the passband for the tolerance-aware optimum and better for the na\"ive optimum. 98 of the 100 random samples in the uncertainty box-set have transmission better than 95\% across the passband. Furthermore, the transmitted power of this transmission extends beyond the passband; the tolerance-aware optimum point design has a half power ratio bandwidth of 1:2.3, neglecting the narrow resonance just above \qty{200}{\GHz} which may be unphysical.
	
	\begin{table}[t]
		\caption{\Acl{WG} to microstrip transition parameter values. The parameters $x^\text{min}_i$ give the solution to na\"ive optimization with $\argmin{f(\mathbf{x})}$, and the parameters $x^\star_i$ give the solution to the tolerance-aware optimization problem described by \cref{eqn:optalg} (left table). The other values (right table) were held constant during this optimization, except for the parameter $t$, which was na\"ively optimized but excluded from the full tolerance-aware optimization to reduce the computational complexity of the convex hull construction. $W$ is the microstrip conductor trace width and $H$ is the device layer thickness. The features associated with these parameters are illustrated in \cref{fig:uStrip_Diagram}.}
		\begin{tabular}[t]{@{} l S[table-format=3.1] S[table-format=3.1] @{}}
			\toprule
			\thead{Param.} & \multicolumn{1}{c}{\thead{$x^\text{min}_i$\,{}[\unit{\micro\meter}]}} & \multicolumn{1}{c}{\thead{$x^\star_i$\,{}[\unit{\micro\meter}]}} \\
			\midrule
			$x_1$    & 285 & 262      \\ %
			$x_2$    & 262 & 278       \\ %
			$x_3$    & 470 & 457      \\ %
			$x_4$    & 559 & 563      \\ %
			$x_5$    & 156 & 184      \\ %
			$x_6$    & 230 & 212      \\ %
			$x_7$    & 198 & 159      \\ %
			$x_8$    & 104  & 105       \\ %
			$x_9$    & 90 & 89       \\ %
			$x_{10}$ & 103 & 108       \\ 
			$x_{11}$ & 30 & 48        \\ 
			$\Delta x$& \multicolumn{1}{c}{---} & 6.2 \\
			\bottomrule
		\end{tabular}
		\hfill
		\begin{tabular}[t]{@{} l S[table-format=3.1] @{}}
			\toprule
			\thead{Param.} & \multicolumn{1}{c}{\thead{Value\,{}[\unit{\micro\meter}]}} \\
			\midrule
			$a$   & 1597              \\ %
			$b$   & 820               \\ %
			$t$   & 287               \\ %
			$w_p$ & 180               \\ 
			$l_m$ & 100               \\ 
			$t_b$ & 164               \\ %
			$w_a$ & 230              \\ 
			$w_t$ & 100                \\ 
			$h_t$ & 213                \\
			$l_h$ & 325                \\
			$W$   & 25.0              \\ %
			$H$   & 25.0              \\
			\bottomrule
		\end{tabular}
		\label{tab:ustrip_params}
	\end{table}
	
	\section{Measured Device} \label{app:measured}
\begin{figure}[t]
		\centering
		\includegraphics[width=\linewidth]{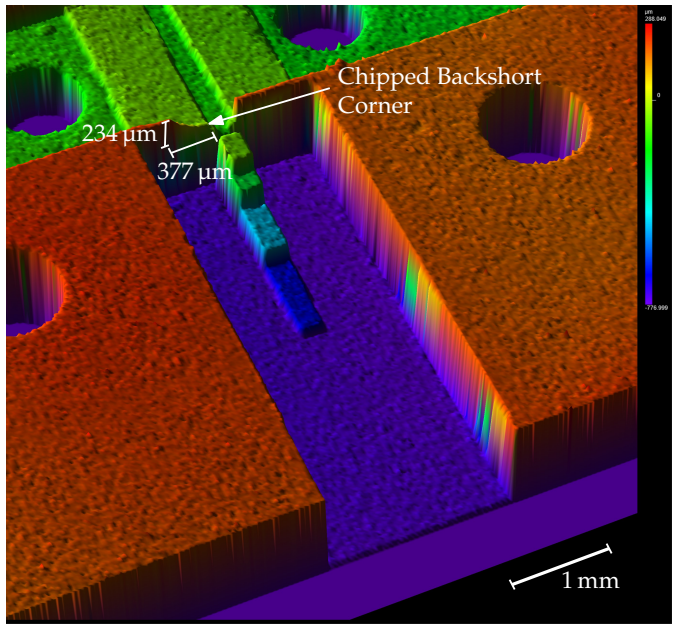}
		\caption{Confocal micrograph of the waveguide structure used for the measured device. The color of the surface corresponds to its height. Note the chipped corner of the backshort, which was included in the warm simulations. We discuss the cause of and solution to this defect in \cref{sec:meas}.}
		\label{fig:measured_device}
	\end{figure}

	\Cref{fig:measured_device} depicts the waveguide structure for the measured device, including the chipped backshort corner defect. 
	
	\section{3D-Printed Surface Characterization} \label{app:surface}
\begin{figure}[t]
		\centering
		\includegraphics[width=\linewidth]{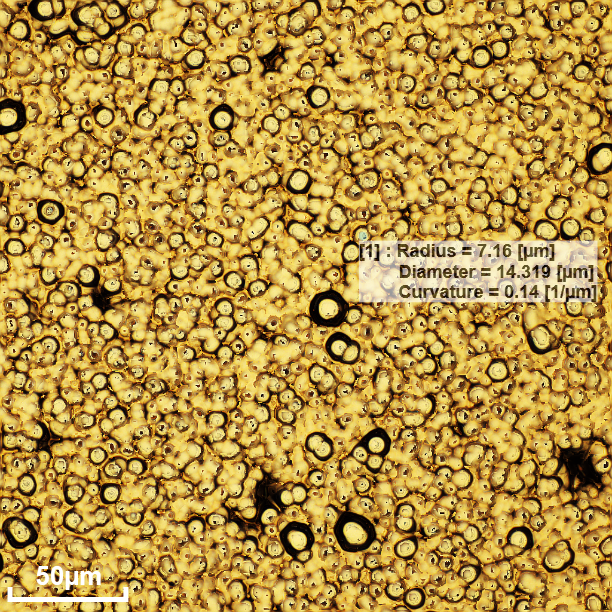}
		\caption{Optical micrograph of a patch of the metallized 3D-printed surface using a $50\times$ magnification objective lens. Note that the surface appears to be a conglomeration of spherical nodules. Confocal microscopy measurements of a typically sized nodule using a spherical fit are shown.}
		\label{fig:surface_sample}
	\end{figure}

	The \ac{WG} portion of the \ac{WG-CPW} transition is metallized after being 3D-printed. The resulting surface appears as a conglomeration of spherical nodules. \Cref{fig:surface_sample} shows an optical micrograph of a typical patch of the surface of a test device which was produced by a confocal microscope that also recorded three-dimensional measurements of the surface. Measurements across this patch give a nodule radius range of \qtyrange{5}{15}{\micro\meter} with a nodule density estimated at \qty{30.5}{\micro\meter/nodule} over a \qty{947}{\micro\meter^2} sample area in this patch. Note that the surface roughness recorded in this patch is significantly smaller than the nodule radius, with $S_a=\text{\qty{0.5}{\micro\meter}}$ and $S_q=\text{\qty{0.6}{\micro\meter}}$, since most of the spherical nodules protrude only a small amount from the surface.
	
	\section*{Backmatter}
	
	\begin{backmatter}
		
		\bmsection{Acknowledgments} %
		
		Thanks to Norman Jarosik of Princeton for transition measurements in the lower half of the frequency band. This work was primarily supported by NASA grant 80NSSC22K1746. This work was supported by a NASA Space Technology Graduate Research Opportunity under NASA grant 80NSSC24K1378. This work made use of the Pritzker Nanofabrication Facility of the Institute for Molecular Engineering at the University of Chicago, which receives support from the Soft and Hybrid Nanotechnology Experimental (SHyNE) Resource (NSF ECCS-2025633), a node of the National Science Foundation’s National Nanotechnology Coordinated Infrastructure. Thanks to the staff of the Pritzker Nanofabrication Facility for their fabrication expertise which helped make this work possible: Will Hyland, Julia Krueger, Jered Feldman, Sam Kaehler, Kendall Davis, and Peter Duda. This work made use of the shared facilities at the University of Chicago Materials Research Science and Engineering Center, supported by the National Science Foundation under award number DMR-2011854. Thanks also to Thomas Cecil of Argonne National Laboratory for his fabrication advice. Furthermore, this document was prepared using the resources of the Fermi National Accelerator Laboratory (Fermilab), a U.S. Department of Energy, Office of Science, Office of High Energy Physics HEP User Facility. Fermilab is managed by Fermi Forward Discovery Group. J.Z. is also supported by the Kavli Institute for Cosmological Physics. This material is based on work supported by the National Science Foundation Graduate Research Fellowship under Grant No. 2140001.

		\bmsection{Disclosures} The authors declare no conflicts of interest.

	\end{backmatter}

	\bibliography{MAIN}%

\end{document}